\documentclass[runningheads]{llncs}
\usepackage{graphicx}
\usepackage{adjustbox}
\usepackage[section]{placeins}
\usepackage{float}

\begin{document}
\title{Cloud Based Big Data DNS Analytics at Turknet}

\author{Altan Cakir\inst{1}\ \and
Yousef Alkhanafseh\inst{2} \and
Esra Karabiyik\inst{3} \and
Erhan Kurubas\inst{4} \and
Rabia Burcu Bunyak\inst{5} \and
Cenk Anil Bahcevan\inst{6}}
\authorrunning{Cakir et al.}

\institute{Istanbul Technical University, Sariyer, 34467, Turkey \\ 
\email{altan.cakir@itu.edu.tr}\\
\and
Turknet Iletisim Hizmetleri, Sisli, 34394, Turkey\\
\email{\{yousef.alkhanafseh,esra.karabiyik,erhan.kurubas,\\burcu.bunyak,cenk.bahcevan\}@turknet.net.tr}}

\maketitle              % typeset the header of the contribution
\begin{abstract}
Domain Name System (DNS) is a hierarchical distributed naming system for computers, services, or any resource connected to the Internet. A DNS resolves queries for URLs into IP addresses for the purpose of locating computer services and devices worldwide. As of now, analytical applications with a vast amount of DNS data are a challenging problem. Clustering the features of domain traffic from a DNS data has given necessity to the need for more sophisticated analytics platforms and tools because of the sensitivity of the data characterization. In this study, a cloud based big data application, based on Apache Spark, on DNS data is proposed, as well as a periodic trend pattern based on traffic to partition numerous domain names and region into separate groups by the characteristics of their query traffic time series. Preliminary experimental results on a Turknet DNS data in daily operations are discussed with business intelligence applications.

\end{abstract}
\section{Introduction}

The Domain Name System (DNS) is part of the core infrastructure of the Internet. Measurement of analtical changes in the DNS traffic over time provides valuable information about the evolution of business intelligence and predictive maintanance operations for internet companies.  We, Turknet, use a complementary approach based on active measurements, which provides a unique, comprehensive dataset on the evolution of DNS over time. Our cloud based high-performance infrastructure platforms and integrated developments performs Internet-scale active measurements, currently offline querying over of the DNS name space on a daily basis. Our infrastructure is designed from the ground up to enable big data analysis approaches on, e.g., a Apache Spark  and Hadoop cluster. With this novel approach we aim for a customer-oriented DNS-based measurement and analysis of the Internet traffic at Turknet.

\section{Analysis Tools}
\subsection{Apache Spark}

Apache Spark is a parallel processing framework that is designed to perform a series of tasks at the same time. It has several distinctive features that make it preferable among other big-data solutions. High processing speed, ability of recovering data, supporting various software development languages, and working in-memory computation model are some examples of Spark traits. Due to its advantages, currently, it is almost used in all data engineering applications such as network, banking, security, manufacturing, media, etc. Commonly, Apache Spark structure consists of eight main components, which are Spark core, Spark SQL, Spark streaming, machine learning libraries, graph processing, and cluster management\cite{ref_book1}. The prime operations of Spark for Big data engineering, including elements for memory management, task scheduling, performance improvement, and storage systems, and other, are located in Spark Core design. Furthermore, spark core contains various APIs that determines immutable objects (Resilient Distributed Datasets (RDDs), Dataframe and DataSets), which can be considered as Spark’s heart components\cite{ref_url1}. Spark SQL is useful with structured data and it permits querying data via SQL for stuructured datasets. On the other hand, analyzing live streams of data can be achieved using spark streaming including on-they-fly machine learning solutions. Lastly, the spark cluster manager, which could be spark’s own standalone, YARN (Yet-Another-Resource-Negotiator), or Mesos, specifies the jobs which will be achieved by the spark cluster. Unlike traditional MapReduce methods, Apache Spark supports both interactive queries and stream processing which gives it huge processing speed while treating large set of data. As a consequence, several minutes or maybe hours will be eliminated from processing data execution time when it is compared to other traditional MapReduce methods in this analysis. Besides, Spark operations and computations are run in memory not in disk which reduces the risk of going disk to full usage while dealing with complex applications \cite{ref_url2}.

\subsection{Processing big data with computing cluster: EMR}
Amazon Elastic MapReduce (EMR) is one of the analytic services that Amazon Web Service (AWS) provides as a platform-as-a-service (PaaS). EMR, basically, hosts Apache Hadoop framework that is entirely built on instance and storage at cloud platform, aka Amazon Elastic Compute Cloud (EC2) and Amazon Simple Storage Service (S3). Thereby, there is no need to deal with the intricacy of installing Apache Hadoop cluster or paying for it.  In general, EMR is used as a distributed computing environment (cluster) that number of its master and core nodes can be scaled up or down on demand without the need to establish complex structure of hardware\cite{ref_whitepaper}. EMR Hourly prices can be considered appropriate when it is compared with its potential force and its prices ranges, such as other cloud vendors\cite{ref_url3}. EMR cluster is able to process several tasks of massive datasets in parallel within few minutes as well.

\section{Data Analyzing Process}
Traditionally, the general data flow diagram in EMR, as shown in \textbf{Fig.1}, starts with moving data to an AWS S3 bucket and ends with estimating the cost and the performance of the process. In between, data collection, aggregation, and process are accomplished.

\begin{figure}
\includegraphics[width=11.5cm,height=1.75cm]{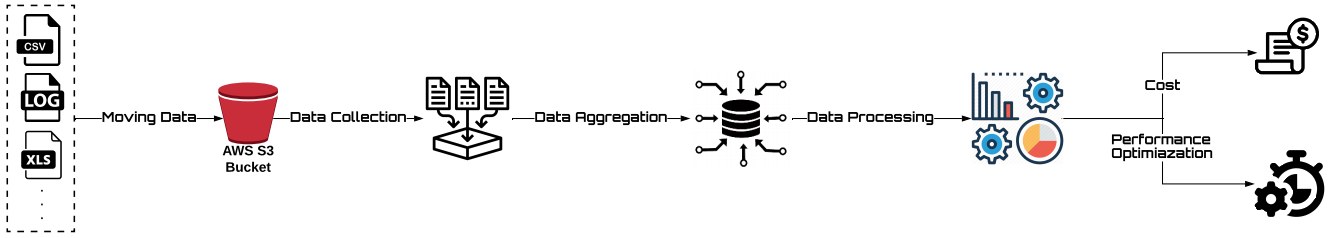}  
\caption{General Data Flow Scheme.}
\end{figure}

\subsection{System Architecture}
The Structural Flow of Turknet DNS Data analysis is represented in \textbf{Fig.2}. This figure starts from the left side with Turknet DNS servers logs. In this figure, the customers are represented as client1, client 2, client 3 and client n. Generally, Turknet company infrastructure contains three different Domain Name Servers (DNSs), which have the ability to collect consumers information, while they are browsing various internet sites. Then, by using specific methods, Turknet user’s data that is located in DNSs can be extracted and saved in log format files. Since the extracted data from DNSs are considered as high volume (15 TB/month), it is not plausible to use either internal nor external hard disks to save this massive volume of data. For example, in order to save 15 TB of data, 4 storage devices that each of which has the ability to store about 4 TB are required, which will likely increase the opportunity of losing data. Fortunately, there is cloud based S3 solution, which helps in overcoming this problem by using its buckets that can manage and process huge size of data. This method could thereby eliminate the need of utilizing several data storage devices in order to store large volume data. However, s3 or similar cloud storage solutions  may lead exorbitant prices, which requires advanced level optimization for the big data operation. Writing and reading data in s3 cloud storage bucket have extra prices as well. This problem could be solved by reducing the size of data which is achieved in Apache Parquet (Apache Spark native) format step. After deploying the log data in AWS s3 bucket, another required data of multiple information related to the users, that is taken from Turknet database, should also be uploaded to the AWS s3 that is just established. Then, the most complex step in this operation appears as the optimization of analyzing the data.

An Amazon Elastic Map-Reduce (EMR) is used in efficient way to complete this step. EMR, in general, is a cluster that owns one or more master nodes and multiple core (slave) nodes. It has several built-in programs such as python programming software language, zeppelin-Apache software, and Apache-spark engine. Coding part is done on Apache Zeppelin integrated development environment (IDE). However, analyzing data step consists of reading, filtering, comparing, and writing these different types of data. The different data is read by using Apache Spark resilient distributed data -sets (RDD) and DataFrame immutable object formats. 
RDD has to be used since the data must be distributed among the cluster slave nodes while processing it. This contributes to a serious reduction of the processing time. The second part of this process is filtering the data. At this part, the useless data-sets such as null, incomplete, and duplicated ones are completely sanitized by using filter and map functions in computing layer. This will end up with the filtered data. After that, the data of DNSs and the data of database are compared with each other by using Structured Query Language (SQL) Dataframe Spark functions such as join and groupBy. The last step in analyzing data process is writing the final data. It is written in Parquet format in partition method by using Spark Dataframe. Parquet format is used solely due to its effectiveness in minimizing the size of the data. Then, parquet files are deployed in another s3 bucket. Finally, the modified data is ready to be visualized by using the numerous types of visualization tools. Two different ways are shown in the Fig.2. below which are Elasticsearch Logstash and Kibana (ELK), and Plotly methods. In ELK, the parquet data can be directly used by loading it to elasticsearch. Then, it should be loaded from elasticsearch to an index in kibana by using Logstash. In Kibana, which is an open source visualization data, the data can be presented in different chart types such as time-lion, bar chart, line chart, etc. The other way is to use Plotly. Using Plotly, only, needs additional step which is converting the data file format from parquet to CSV format.

\begin{figure}
\includegraphics[width=\textwidth,height=5cm]{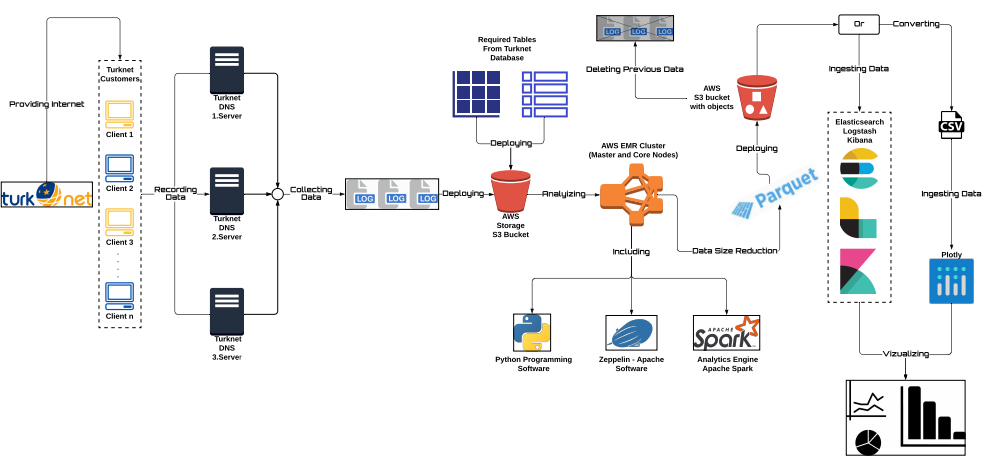}
\caption{Turknet DNS Analytics System Architecture.}
\end{figure}

Several experiences were examined on EMR using three different types of cluster instances (master and node cores) in order to obtain the most optimal configuration, and hence, considering it as the typical one in the forward operations. The tested instances are m5.xlarge, m5.2xlarge, and r5.4xlarge which their major features of vCPU and RAM can be found from Table1.

\begin{table}
\caption{Features of Three Different AWS EMR Instances.}
  \centering

\begin{tabular}[H]{|c|c|c|c|c|c|}

\hline
\vtop{\hbox{\strut Amazon}\hbox{\strut Node Type}}&vCPUs&RAM&\vtop{\hbox{\strut Instance }\hbox{\strut Storage}\hbox{\strut (GiB)}}&\vtop{\hbox{\strut Network }\hbox{\strut Performance }\hbox{\strut (Gbps)}}&\vtop{\hbox{\strut EBS }\hbox{\strut Bandwidth}\hbox{\strut (Mbps)}}\\
\hline
m5.xlarge & 2 & 8G & EBS-Only & Up to 10 & Up to 4,750 \\ 
m5.2xlarge & 8 & 32G & EBS-Only & Up to 10 & Up to 4,750\\ 
r5.4xlarge & 16 & 128G & EBS-Only & Up to 10 & 4,750\\ 
\hline
\end{tabular}
\end{table}

The target of the process is to make the combination of unstructured and structured data between different data sets and data reduction with increasing the value. As is pointed before, Turknet has three DNS servers which data can be collected from, in this procedure, about 300 GB one day data is captured from DNS servers and analyzed by using Spark application on EMR built-in Zeppelin. Furthermore, the distributed computing cluster settings of Apache Spark application should be configured cautiously, depending on the number of core instances and their virtual CPU cores and their memory, while creating EMR cluster. For instance, the spark application configurations of an EMR cluster of one master and 10 core nodes of r5.4xlarge instance, each node has 16 vCPUs and 128 GiB memory, should be modified as in the following:  dynamicAllocation.enabled - False, executor.cores - 5, executor.memory - 37 GB, executor.instances - 170,and yarn.executor.memoryOverhead - 5 GB. These values of spark application have been calculated based on an article published in AWS official internet site\cite{ref_url4}. The first property which is spark.executor.cores represents the maximum number of tasks that can be executed in parallel in one executor. Next, the heap size, amount of allocated memory, of each executor is controlled by spark.executor.memory. Similarly, spark.executor.instances determines the number of executors that will be run in a spark job. ultimately, with changing spark.yarn.executor.memoryOverhead property the memory requirement to YARN for each executor can be regulated.  As a consequence, the number of parallel tasks that will be run at the same time is equal to the result of multiplying spark.executor.cores with spark.executor.instances.

This operation encourages the Apache spark application to utilize the clusters’ maximum potential ability for its purpose. However, the Apache spark default configurations maybe used as they are for modest applications. After adjusting the proper settings of Apache spark, the process can be started. The first case included 1+10 m5.xlarge EMR cluster. In this case, the python script was run and completed its all tasks within 40 minutes. Despite its high running time it is financial appropriate with 0.352 dollar for the whole process.  The other case included the same number of instances but with considering more vCPU cores and bigger memory size. The elapsed time of this case was approximately 22 minutes which fulfills a 16 minutes reduction and a little increase in billing. Finally, another type of AWS instances was employed which is r5.4xlarge. This type ac hived astonishing reduction in processing time. The elapsed time of this process was approximately 13 minutes. However, comparing r5.4xlarge with m5.xlarge leads to save nearly 27 minutes. It can be said, as a outcome, that the difference in price (which is 0.2037 per Hour) between m5.xlarge and r5.4xlarge can be significantly compensated since the running time of r5.4xlarge is less than the one of m5.xlarge (see Table 2).

\begin{table}
\caption{Three Different EMR Cluster Study Cases.}
  \centering

\begin{tabular}{|c|c|c|c|c|c|c|}
\hline
\vtop{\hbox{\strut Amazon Node}\hbox{\strut Type}}&\vtop{\hbox{\strut Cluster Size}}&\vtop{\hbox{\strut Processed Data}\hbox{\strut Size (GB)}}&\vtop{\hbox{\strut Output Data}\hbox{\strut Size (GB)}}&\vtop{\hbox{\strut Running Time}\hbox{\strut (minutes)}}\\
\hline
m5.xlarge&&&&40\\
m5.2xlarge&1 + 10&246&29&22\\
r5.4xlarge& && &13\\
\hline
\end{tabular}
\end{table}

\section{Results}

The analysis main measurement runs on a cloud-based EMR cluster. Every DNS log traffic files with CRM and CDR files are orchestrated by a cluster management system architecture at AWS. This architecture is responsible for distributing chunks of work, of 3-4 billion rows including domains each, to a set of worker nodes. Master node runs custom-built configuration on slave nodes that performs a pre-defined selection of DNS queries for each domain in a chunk of work. Queries are matched against a instance running on the worker node. Data analyzed by workers is sent to a central aggregation point s3 and visualization for each user for further processing and analysis. As case study, we have analysed the use of over one week period. We focused on interesting particular domain categorization with respect to internet traffic. Fig.3. shows the growth in the fraction of domains consumption per server that use either of these platforms. Growth is presented as a number of unique household customer relative to the start of the 7-day period. To perform the analysis, our platform processed over 25 billion query results. The full analysis was performed by 11 r5.4xlarge node in about 13 minutes for each day. This can easily be improved by running the analysis on a larger cluster configuration. This example shows what can be achieved using our measurement platform and data. The growth in use of DNS servers illustrates how the Internet usage is evolving from every organisation managing its own services, including predictive maintenance, to a few large providers offering these services in future.

The figure (Fig.3.) below presents the hourly changes of active users on the data collected from different DNS of Turknet. Graph shows the active user decrease on 4 am. to 6 am. and the increase on 8 pm. to 12 pm. On the other hand, The heat map of DNS network traffic of Turkey can be clearly showun in Fig 4.

\begin{figure}[H]
    \includegraphics[width=11.5cm,height=9.5cm]{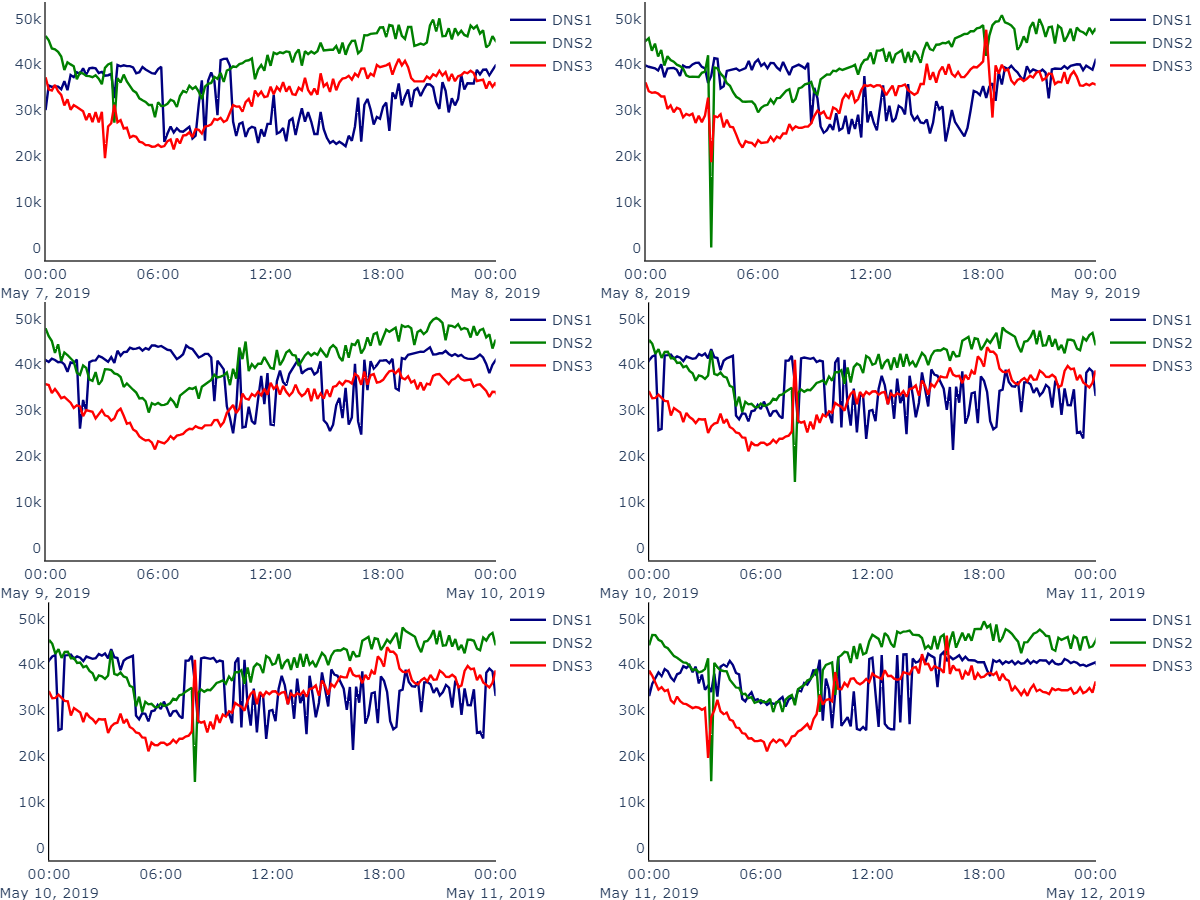}
    \caption{Active unique user numbers between 7 and 13 May 2019 on Turknet network.}
\end{figure}

\begin{figure}[H]
    \centering
    \includegraphics[width=12cm,height=3.5cm]{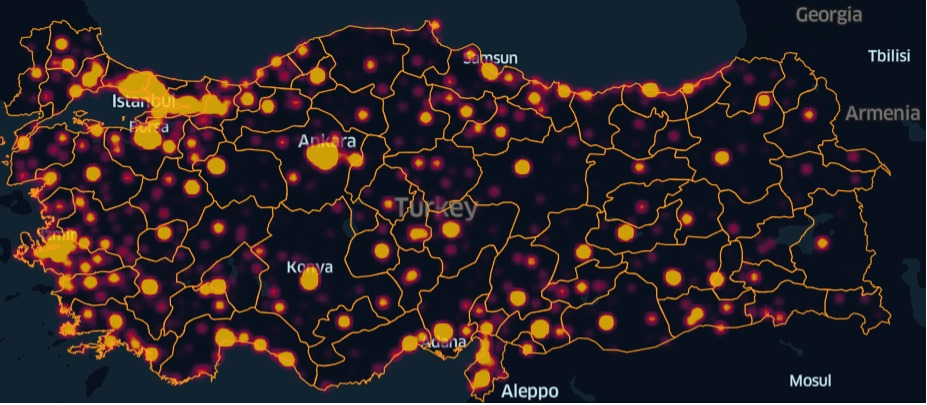}
    \caption{Density Map of 30 May 2019 of URL Categories Distribution.}
\end{figure}

Lastly, the categories of Uniform Resource Locator(URL) are obtained and analyzed. Several categories can be handled. However, as it is obvious from Fig 5., six categories are neatly considered. Technology/Internet category which is represented in red, was the highest one among the other categories with nearly 74 thousand of unique usage at 00:00 o'clock .

\begin{figure}[H]
    \centering
    \includegraphics[width=11.5cm,height=4cm]{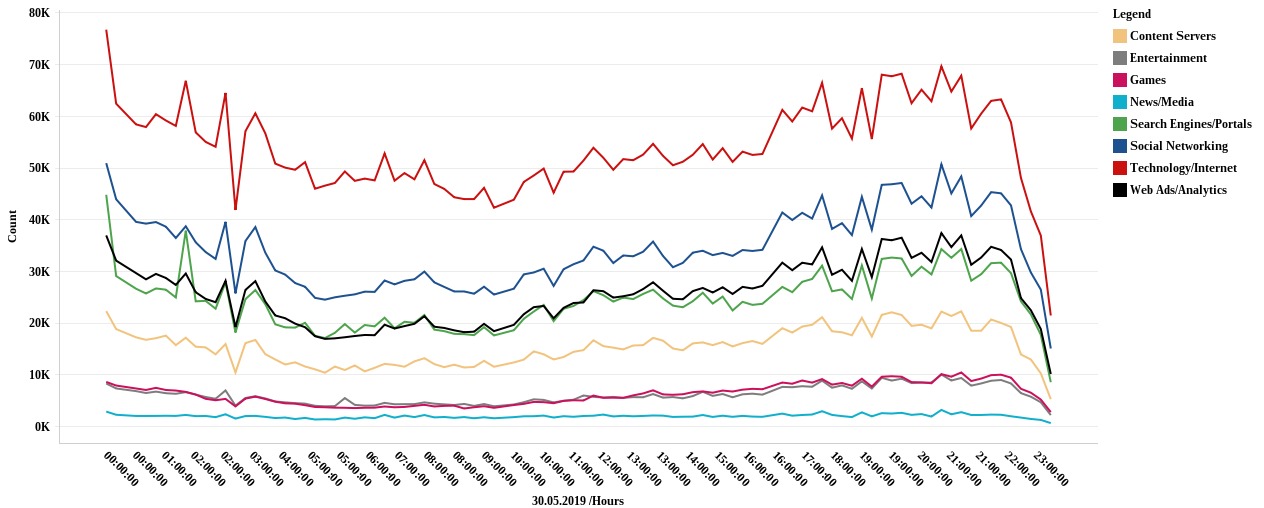}
    \caption{Traffic of 30 may 2019 of URL Categories Distribution Over Time.}
\end{figure}

\section{Conclusion}
We designed a unique active measurement infrastructure for the DNS data with other various data-sets. Our infrastructure actively measures the total DNS traffic on a daily basis. The resulting output enables reliable DNS-based analysis of the evolution of the Internet for the first time at Turknet. And not only do we measure on a large scale, we have also carefully designed for optimal analysis of the collected data through the Apache Spark cluster optimization in EMR tool-chain. The simple case study included in this work showcases use of our data-set. It answers the simple question about the tremendous scale of DNS traffic data can be tractable and analyzed in an effective implementation of a cloud based big data platform. This provides a cost effective business intelligence application for the big data analytic at scale in Turknet.


\end{document}